\documentclass[useAMS,usenatbib]{mn2e}
\usepackage{epsfig}
\usepackage{subfig}
\usepackage{times}
\usepackage{color}

\newcommand{\teff}{$T_{e\!f\!f}$}
\newcommand{\kms}{km s$^{-1}$}
\newcommand{\logg}{$\log g$} 
\newcommand\rgc{$R_{\rm G}$}
\newcommand\vgc{$V_{\rm GC}$}

\title[ARGOS bulge survey]{ARGOS IV: The Kinematics of the Milky Way Bulge}
\author[M. Ness et al.]
  {M.~Ness$^1,$\thanks{E--mail:mkness@mso.anu.edu.au.}
  K.~Freeman$^1$,  E.~Athanassoula$^2$, E.~Wylie--de--Boer$^1$, 
  J.~Bland--Hawthorn$^3$, 
  \newauthor 
 M.~Asplund$^1$,
 G.F~Lewis$^3$,
  D.~Yong$^1$,
 R.R.~Lane$^4$,
  L.L~Kiss$^{3,5,7}$,
  and R.~Ibata,$^6$  \\
$^1$Research School of Astronomy \& Astrophysics, Australian National University, Cotter Rd., Weston, ACT 2611, Australia\\
$^2$Aix Marseille Universit\'e, CNRS, LAM (Laboratoire d'Astrophysique de
Marseille) UMR 7326, 13388, Marseille, France  \\
$^3$Sydney Institute for Astronomy, University of Sydney, School of Physics A28, NSW 2006, Australia\\
$^4$Departamento de Astronom\'{i}a Universidad de Concepci\'{o}n, Casilla 160 C, Concepci\'{o}n, Chile\\ 
$^5$MTA CSFK, Konkoly Observatory, Konkoly Th. M. 15-17, H-1121, Hungary. \\ 
$^6$Observatoire Astronomique, Universit\'{e} de Strasbourg, CNRS, 11 rue de l'Universit\'{e}, F--67000,      Strasbourg, France \\
$^7$ELTE Gothard-Lend\"ulet Research Group, H-9700, Szombathely, Hungary.\\
 } 

\begin{document}

\date{Revision A sent to ARGOS team on 3 Dec 2012}

\pagerange{\pageref{firstpage}--\pageref{lastpage}} \pubyear{2011}

\maketitle

\label{firstpage}

\begin{abstract}

We present the kinematic results from our ARGOS spectroscopic survey of the Galactic bulge of the Milky Way. Our aim is to understand the formation of the Galactic bulge. We examine the kinematics of about 17,400 stars in the bulge located within 3.5 kpc of the Galactic centre, identified from the 28,000-star ARGOS survey. We aim to determine if the formation of the bulge has been internally driven from disk instabilities as suggested by its boxy shape, or if mergers have played a significant role as expected from $\Lambda$CDM simulations. From our velocity measurements across latitudes $b =$ --5$^ \circ$, --7.5$^\circ$ and --10$^\circ$ we find the bulge to be a cylindrically rotating system that transitions smoothly out into the disk. Within the bulge, we find a kinematically distinct metal-poor population ([Fe/H ]  $< -1.0$) that is not rotating cylindrically. The $5\%$ of our stars with [Fe/H]  $< -1.0$ are a slowly rotating spheroidal population, which we believe are stars of the metal-weak thick disk and halo which presently lie in the inner Galaxy.   The kinematics of the two bulge components that we identified in ARGOS paper III (mean [Fe/H] $\approx$ --0.25 and  [Fe/H] $\approx$ +0.15, respectively) demonstrate that they are likely to share a common formation origin and are distinct from the more metal-poor populations of the thick disk and halo which are co-located inside the bulge. We do not exclude an underlying merger generated bulge component but our results favour bulge formation from instabilities in the early thin disk. 

\end{abstract}


\section{Introduction}

The principal goal of the ARGOS spectroscopic survey is to obtain chemical and kinematic data for a substantial number of stars to understand the origin of the Galactic bulge of the Milky Way. The two main scenarios for bulge formation are hierarchical merging as seen in Lambda Cold Dark Matter ($\Lambda$CDM) simulations of Galaxy formation \citep{Abadi2003a} and disk instability \citep{Combes1981, Raha1991, Athanassoula2005}. These different formation processes imprint distinct kinematic signatures and thus can be tested using observations of the stellar kinematics of the bulge. %

Boxy/peanut bulges rotate almost as rapidly at high latitude as in the plane (cylindrical rotation). This has been seen in observations of boxy bulges in other galaxies, e.g. \citet{FalconBarroso2006}, and in numerical simulations \citep{Athanassoula2002a}. This is unlike the merger-generated `classical' bulges which show slow rotation at high latitudes. \citet{Howard2009} have identified such cylindrical rotation in the Galactic bulge for a subsample of the more metal-rich stars. Shen et al. (2010) constrain the bulge to have no underlying merger-generated component to a level of $\le$ 8$\%$.

The ARGOS survey sample is unbiased in metallicity, and provides kinematic data of stars in the Galactic bulge over the wide range of abundance that is a consequence of the bulge formation process.   This may enable us to investigate at what stage the boxy bulge structure was formed during the evolution of the inner Galaxy. 

We test the prediction of rapid high-latitude rotation by measuring velocities for bulge stars in 25 fields at three Galactic latitudes in the south (plus three fields in the northern bulge as checks on symmetry), out into the disk across a longitude range from $+26^\circ$ to $-31^\circ$. Our sample size is large enough to detect a $5\%$ contribution from a classical slowly-rotating merger-generated spheroidal component underlying the dominant boxy/peanut bulge.  We note, however, that \citet{Saha2012} have shown that a weak classical bulge would be spun up into rapid rotation by the boxy/peanut bulge, and would therefore be difficult to detect kinematically. The ARGOS survey would also include stars of the metal-poor ``first stars'' population, if they are present in the inner bulge as predicted by cosmological simulations \citep{Diemand2005}.

This paper first reintroduces, in Section 2, the stellar populations of the bulge identified in ARGOS paper III \citep{Ness2013}. Sections 3-5 briefly recapitulate the details of the observations, stellar sample selection and the methods used to measure stellar parameters and determine distances. The distance determination has enabled  selection of a relatively uncontaminated sample of stars in the inner Galaxy.  Using our estimates of stellar parameters,  we have been able to identify and eliminate foreground and background stars.  In Section 6, we present our results on the cylindrical rotation profile obtained for the bulge and the slow rotation of the metal-poor population with [Fe/H] $<$ --1.0 in the inner region of the Galaxy. We also compare the kinematics of populations of different metallicity and provide  evidence for a common dynamical history for stars with  [Fe/H] $> -0.5$. In Section 7, we compare our kinematics with those for N-body models and discuss the origin of the components we find in the bulge in Section 8.  Conclusions are presented in Section 9. 

\section{Stellar Populations in the Bulge}

\begin{figure}
\centering
    \includegraphics[scale=0.17]{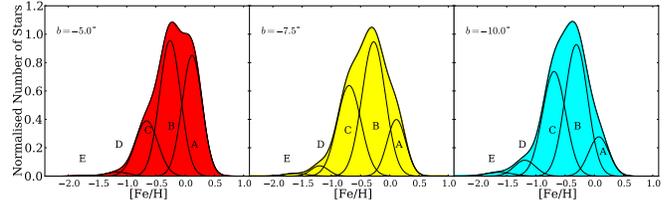}
\caption{MDFs for stars within \rgc\ $<$ 3.5 kpc at from left to right $b=-5^\circ$, $b=-7.5^\circ$ and $b=-10^\circ$, for $l= \pm 15^\circ$, showing the changing contribution of metallicity fractions with latitude.}
\label{fig:components}
\end{figure}

ARGOS paper III \citep{Ness2013} presents the metallicity distribution of the bulge.  We find that the bulge is a composite population. Figure \ref{fig:components} shows our decomposition of the ARGOS metallicity distribution functions (MDFs) at $b = -5^\circ$, $-7.5^\circ$ and $-10^\circ$ for stars in the bulge with \rgc\ $<$ 3.5 kpc across $l=\pm 15^\circ$, into five Gaussian components. These components are labeled A-E in order of decreasing [Fe/H]. We used Gaussians for simplicity in this analysis and the number of components fit was motivated by the distinct peaks seen in the metallicity distribution function and a statistical analysis; components A-C are our primary components and D and E are included to represent the small fraction of very metal poor stars in the sample which increase in fraction further away from the Galactic plane \citep[see][for a discussion]{Ness2013}. These components are proxies for stellar populations and are used as a tool to measure the changing contribution fraction of stars with different metallicities as a function of ($l,b$). The analysis of the metallicity-dependent split red clump stars along the minor axis fields \citep{Ness2012} has demonstrated that stars with [Fe/H] $>$ --0.5 are part of the boxy/peanut bulge and stars with [Fe/H] $<$ --0.5 are not. 

If the formation of the bulge of our Galaxy is indeed a result of a dynamical instability of the initial disk \citep{Athanassoula2005, Shen2010} then it is critical to interpret and analyse the bulge in terms of the contribution of these stellar components which reflect the way in which the bulge was formed from the pre-existing disk. For the bulge stars with [Fe/H] $>$ --0.5, there are two populations (components A and B). Component A is a metal-rich, relatively thin boxy/peanut-bulge population with a mean [Fe/H] = +0.15 and is concentrated close to the plane. We associate this component  chemically with the thin disk in the inner region. Component B is a more metal-poor (mean [Fe/H] = --0.25) and thicker boxy/peanut-bulge structure, distributed across $b=-5^\circ$ to $b=-10^\circ$.  We have argued in ARGOS III that component B represents the stars formed out of the thin disk at early times. 

In summary, contributing to the stellar density in the bulge region are the thin boxy/peanut-bulge (A), the thick boxy/peanut-bulge (B), the thick disk (C), metal-weak thick disk (D) and halo (E), in order of decreasing [Fe/H].  The stars of the halo are probably not part of the bulge, in that they likely have a different origin from the majority of the stars in the bulge. These halo stars are likely on highly eccentric orbits passing through this region at this time. 

\section{Observations and Strategy}

\begin{figure}
\centering
  \includegraphics[scale=0.28]{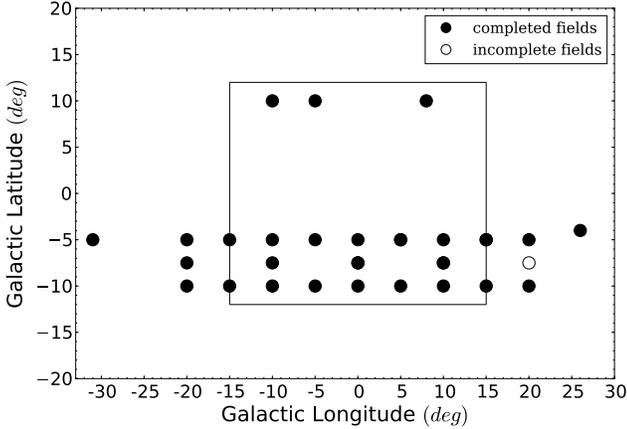}
 \caption{Galactic latitude and longitude for the 28 two-degree-diameter fields in our survey. The filled circles indicate fields that were completely observed.  We are missing stars only in one field, at $(l,b) = (20^\circ, -7.5^\circ)$ for which only 600 stars were observed (rather than 1000 typically per field). The inner rectangle represents the bulge region.}
\label{fig:donefields}
\end{figure}

We acquired our spectra with the fiber-fed AAOmega system on the Anglo Australian Telescope. Figure \ref{fig:donefields} shows the locations of the 28 fields, each of two-degree diameter, selected for this program. We observed about 28,000 stars between 2008 and 2011. Our survey covers the bulge, which is estimated from the 2MASS stellar profile to cover a longitude range of  $\pm 15^\circ$ and 
extend in latitude to about $\pm12^\circ$ \citep{Dwek1995}. Our fields extend in longitude out into the thin and thick disk, in order to measure the stellar kinematics at the transition of the bulge and disks.

Each field contains about 1000 stars and a full observation at each $(l,b)$  requires 3 separate setups of the AAOmega fibres. About 350 stars can be observed simultaneously, and each field of 1000 stars was split into three magnitude intervals, as a partial proxy for distance along the line of sight through the bulge. In our setup we used 25 sky fibres for sky subtraction, and 8 fiducial fibres for tracking alignment.

The stellar parameters were derived from a $\chi^2$ comparison with synthetic spectra generated in the 1D Local Thermal Equilibrium (LTE) stellar synthesis program MOOG \citep{Sneden1973} using the Castelli/Kurucz model atmospheres \citep{Castelli2004}. From the distance estimates for our stars, which are based on the stellar parameters, we have isolated the 17,400 stars in our sample within $y  =  \pm 3.5$ kpc (here $y$ is a Cartesian coordinate along the centre-Sun line, with origin at the Galactic centre). 

The full details of our survey, the observations and the analysis, are provided in ARGOS II \citep{Freeman2012}. The metallicity decomposition of populations are discussed in ARGOS III \citep{Ness2013}.

\section{Measuring Radial Velocities}

The radial velocities of stars in our fields were measured from our medium resolution spectra ($R = 11,000$) covering the spectral region $8000 - 8800$\AA\ which includes the strong Ca II infrared triplet absorption lines. Following the data reduction and sky subtraction,  the data were cleaned using a sigma clipping algorithm to remove residual sky lines which improved the accuracy of radial velocity measurements. Radial velocities were obtained via cross correlation of each observed spectrum with a number of template stars generated from synthetic spectral models.

These stellar parameters of the template stars covered a wide range of [Fe/H] at a fixed gravity and temperature. The metallicity values used were [Fe/H] = $-2.0$, $-1.0$ and $0.0$, for \teff \ $= 5000$K and \logg $= 2.8$ (as we expect most of our stars to be clump giants \citep{Zhao2001}). The best matching template was chosen for each  star to measure the radial velocity, as determined by the smallest error of the fit in the cross correlation. The error at this resolution and signal to noise (about 50 to 80) is $\approx$ 0.9 \kms\ (see ARGOS II).  All heliocentric radial velocities were transformed to Galactocentric velocities (see Section 5). 

\section{Bulge Membership}

From the ARGOS spectra, we have measured radial velocities, stellar parameters ( \teff, \logg) and chemical abundance data ([Fe/H], [$\alpha$/Fe]) for our stars. The stellar parameters allow us to identify foreground dwarf stars and, from the estimated distance, to determine which stars are foreground and background giants.

The rotation and dispersion results (see Section 6) are for stars chosen to lie within $y  =  \pm 3.5$ kpc where $y$ is a Cartesian coordinate along the centre-Sun line, with origin at the Galactic centre).

This selection on $y$ is made, rather than a cut in Galactocentric radius \rgc\ $\le$ 3.5 kpc, so as to include the stars in our fields at Galactic longitudes $|l| > 26^\circ$, which have a minimum galactocentric radius, $R_G  > 3.5$ kpc. In total 17,400 stars remain after this distance cut, about 70\% percent of the original sample. 
The remaining 30\% of the stars are in the foreground or background.  For studying the properties of the bulge, it is important to remove these contaminants.

\section{Results}
    
\subsection{Cylindrical rotation}

\begin{figure*}
  \centering
 \epsfig{file=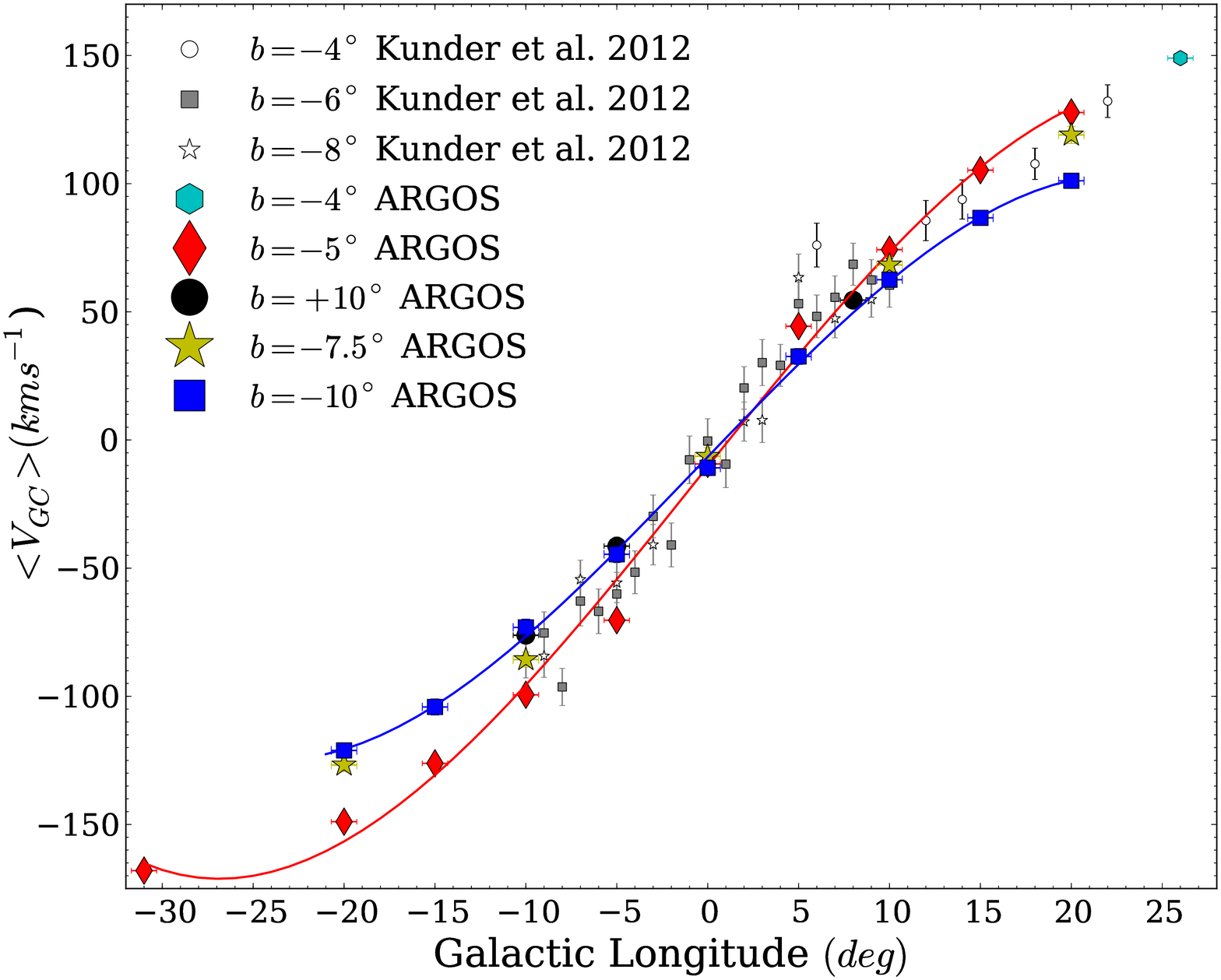,width=0.45\linewidth,clip=} 
 \epsfig{file=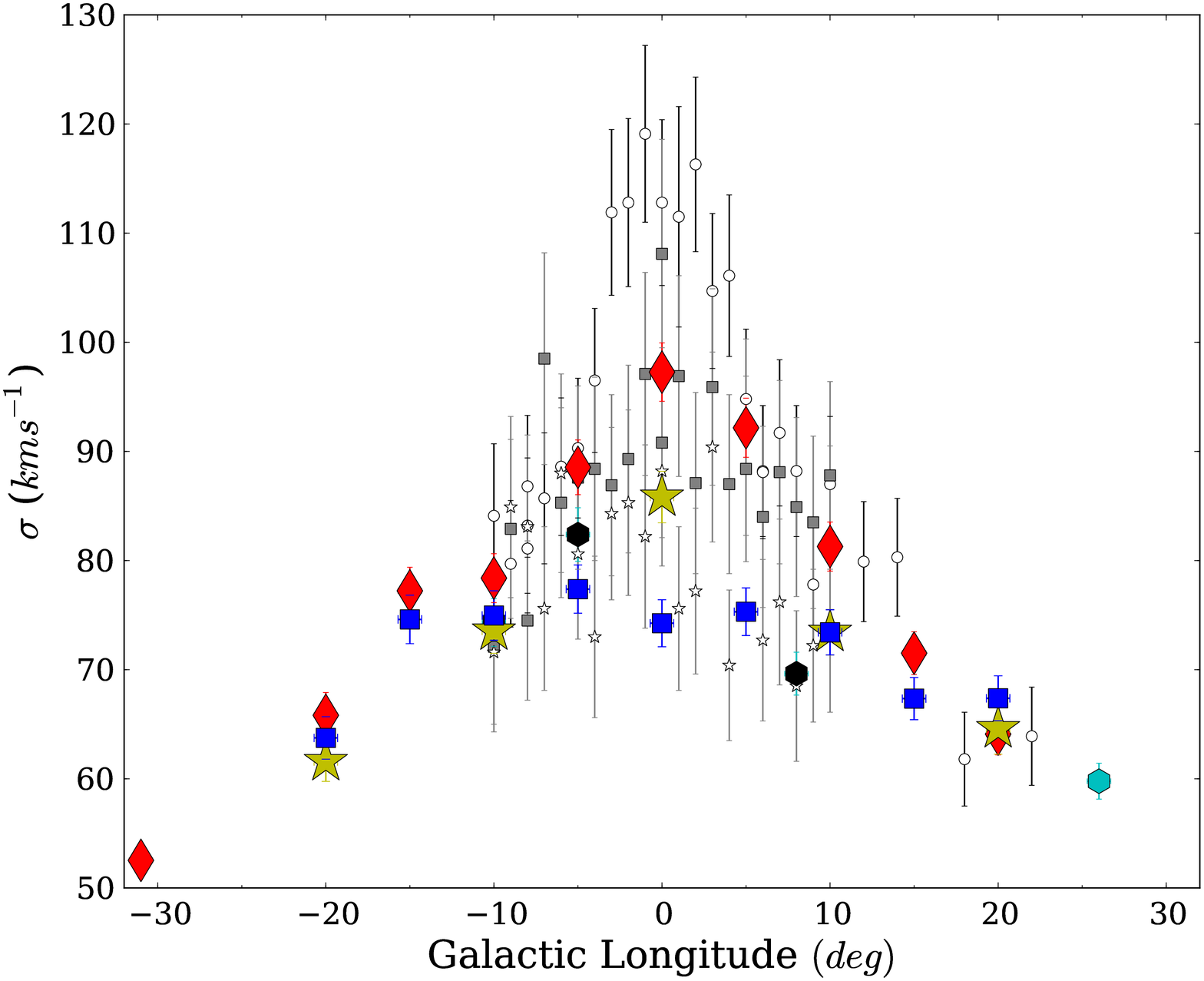,width=0.45\linewidth,clip=} 
   \caption{Rotation curves (left) and velocity dispersions (right) for the 16,600 stars in our survey in the bulge region at y = $\pm$ 3.5 kpc from the Galactic centre with [Fe/H] $>$ -1.0.  The red diamonds are $b=-5^\circ$, yellow stars are $b=-7.5^\circ$, blue rectangles are $b=-10^\circ$ and black squares are $b=+10^\circ$ ARGOS fields. The two curves correspond to our data at $b=-5^\circ$ and $b=-10^\circ$. The horizontal error bars represent the two-degree size of the fields and the vertical error bars are the sampling errors. The BRAVA data \citep{Kunder2012} are also shown using the smaller symbols which correspond to their fields at $b=-4^\circ$ ,$b=-6^\circ$  and $b=8^\circ$ . }
  \label{fig:rotation}
\end{figure*}

Our rotation curves are given in the left panel of Figure \ref{fig:rotation}, and show the Galactocentric rotation velocity in all of our fields out to longitudes of $-30^\circ$ to $+26^\circ$, made using a cut of $|y| < 3.5$ kpc, for stars with [Fe/H] $> -1.0$.  We find cylindrical rotation for the bulge, with only a weak dependence of rotation speed on latitude. These data extend the cylindrical rotation found in the BRAVA survey \citep{Howard2009, Kunder2012} to longitudes of $|l|=10^\circ$. The BRAVA data \citep{Kunder2012} are included for comparison in Figure \ref{fig:rotation} at their latitudes of $b=-4^\circ, -6^\circ$ and $-8^\circ$. 

We corrected for the solar reflex motion, adopting the local standard of rest velocity at the Sun to be 220 \kms\ \citep{Kerr1986} and a solar peculiar velocity of 16.5 \kms\, in the direction $(l,b) = (53^\circ,25^\circ)$  \citep{Binney1981}. The corrected velocity, the Galactocentric velocity \vgc\, is then \\

\noindent \vgc\  =  $V_{HC} + 220[\sin(l)\cos(b)] + 16.5[\sin(b)\sin(25) + \cos(b)\cos(25)\cos(l - 53)]$ \\

\noindent where $V_{HC}$ is the heliocentric radial velocity in km s$^{-1}$ and angles (l,b) are in degrees.

Each data point in Figure \ref{fig:rotation} represents the mean velocity of about 600 ARGOS stars.  The error bars show the sampling error in each field, but due to the large size of our sample they are, in most cases, smaller than the symbol size. Note that the mean velocities of our fields along the minor axis is non-zero: the mean velocities for our three minor axis fields are between about --6 \kms\ to --11 \kms. These non-zero values may come from streaming motions in the bulge, or our adoption of a Sun-centre distance of 8 kpc. Alternatively the offsets could be a sampling effect derived from preferential sampling of more metal-rich stars on the near side of the bulge, which are closer to the plane. Section 6.3 demonstrates that the metal rich stars have lower mean velocities in our fields than the metal poor stars across our longitude range. The rotation is faster on the far side of the bulge at negative longitudes. This may come about because, on the far side of the bulge,
we are observing further along the bulge/bar at a given given longitude, relative to the near side. Systematic distance errors may also contribute.
The rotation increases to $>$ 150 \kms\ at our largest longitudes, which are in the Galactic disk at $l = -31^\circ$ and $+26^\circ$. 

The results for our latitudes of $b=-5^\circ,-7.5^\circ$, $b=-10^\circ$ and also a few fields at $-4^\circ$ and $+10^\circ$ are tabulated in Table \ref{table:velpoints}.  Although we do not share any common fields, our velocity results compare fairly well with those from the BRAVA survey of M giants in the bulge \citep[see][]{Howard2009, Kunder2012}, from $b = -4^\circ$ to $b=-8^\circ$ and mostly across $l = \pm 10^\circ$.  As the lowest latitudes of the BRAVA fields are closer to the Galactic plane, we expect (and confirm) a slightly faster rotation for the stars in their $b=-4^\circ$ field compared to our lowest latitude fields $b=-5^\circ$. The velocities at positive longitudes at $b =-4^\circ$ from BRAVA are higher than the ARGOS measurements at $b=-5^\circ$ which are better matched by the $=-6^\circ$ BRAVA fields.  The rotation and dispersion shows symmetry about the major axis, as seen by comparing the fields at negative and positive latitudes. 

\begin{figure}
  \centering
  \includegraphics[scale=0.32]{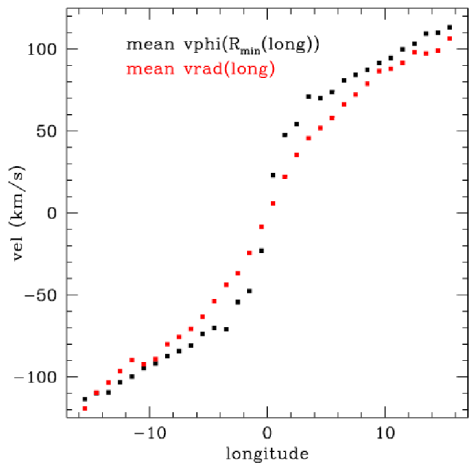}   
   \caption{ The two versions of the rotation curve for the model at $b=-5^\circ$ as a function of longitude.  The red points are calculated as the mean \vgc\ as for the data in Figure \ref{fig:rotation}. The black points show the azimuthally averaged  $V_\phi$ component of the stellar velocity at  the minimum value of \rgc\ at each longitude, i.e. the  true rotation. The red points from the model are for comparison to the data at $b=-5^\circ$ in Figure \ref{fig:rotation}.}
  \label{fig:vrot}
\end{figure}

The rotation curve shown in Figure \ref{fig:rotation} is simply the average value of \vgc\ as a function of longitude, uncorrected for projection.  In this representation, of rotation as a function of longitude, the true rotation curve for the Galaxy would be the mean value for the azimuthal component $V_\phi$ of the stellar velocity for stars with an \rgc\ value close to the minimum \rgc\ at a given longitude. We can use our N-body model as a guide to the difference. Figure \ref{fig:vrot} shows the two rotation curves as derived for the model that we used in \citet{Ness2012}. The red points show the mean \vgc\ which corresponds to the curve shown in Figure \ref{fig:rotation} at a latitude of $b=-5^\circ$. The black points in Figure \ref{fig:vrot} show the mean $V_\phi$ as a function of longitude. At $|l|$ $>$ $10^\circ$ the difference between the mean \vgc\ and the mean $V_\phi$ is only a few percent, but in the inner regions the $V_\phi$ curve is steeper. The red points in Figure \ref{fig:vrot} represent the data well at $b=-5^\circ$ in Figure \ref{fig:rotation}.

\begin{table}
\centering
\begin{tabular}{l  p{1cm}  l  p{0.5cm}  | p{0.5cm}  |  p{1cm}  |  p{1cm}  | p{1cm} |}
\hline \hline
$l$  &  \vgc\  & $V_{err} $ & $\sigma $ & $\sigma_{err} $ & Number of Stars\\
($^\circ$)  & $kms^{-1}$ & $ kms^{-1}$ & $ kms^{-1}$ & $kms^{-1}$ & \\
\hline
\multicolumn{6}{|c|}{$b=-4^\circ$} \\
\hline 
26 & 149.0 & 2.3 & 59.8 & 1.6 &  662 \\
\hline 
\multicolumn{6}{|c|}{$b=+10^\circ$} \\
\hline 
--10 & --75.4 & 3.1 & 74.8 & 2.1 &  587 \\
--5 & --41.4 & 3.5 & 82.4 & 2.4 &  563 \\
8 & 54.5 & 2.8 & 69.9 & 2.0 &  629 \\
\hline
\multicolumn{6}{|c|}{$b=-5^\circ$} \\
\hline 
--31.0 & --168.1 & 2.2 & 52.9 & 1.5 &  597 \\
--20.0 & --149.1 & 3.0 & 65.9 & 2.1 &  498 \\
--15.0 & --126.3 & 3.1 & 77.3 & 2.2 &  639 \\
--10.0 & --99.7 & 3.2 & 78.5 & 2.2 &  614 \\
--5.0 & --70.2 & 3.6 & 88.5 & 2.5 &  622 \\
0.0 & --9.7 & 3.8 & 97.7 & 2.7 &  660 \\ 
5.0 & 44.4 & 3.8 & 92.2 & 2.7 &  579 \\ 
10.0 & 74.2 & 3.2 & 81.3 & 2.2 &  655 \\ 
15.0 & 105.3 & 2.8 & 71.5 & 2.0 &  669 \\ 
20.0 & 127.8 & 2.7 & 64.1 & 1.9 &  564 \\
\hline
\multicolumn{6}{|c|}{$b=-7.5^\circ$} \\
\hline
--20.0 & --127.6 & 2.5 & 60.6 & 1.8 &  568 \\
--10.0 & --85.6 & 2.9 & 73.6 & 2.0 &  645 \\
0.0 & --6.2 & 3.3 & 85.8 & 2.4 &  662 \\
10.0 & 67.9 & 3.0 & 73.4 & 2.1 &  620 \\
20.0 & 118.8 & 3.3 & 64.6 & 2.3 &  380 \\
\hline
\multicolumn{6}{|c|}{$b=-10^\circ$} \\
\hline
--20.0 & --120.4 & 2.7 & 64.1 & 1.9 &  551 \\
--15.0 & --104.2 & 3.1 & 74.6 & 2.2 &  565 \\
--10.0 & --73.1 & 3.2 & 75.0 & 2.3 &  547 \\
--5.0 & --44.8 & 3.1 & 77.4 & 2.2 &  614 \\
0.0 & --10.9 & 3.1 & 74.3 & 2.2 &  592 \\
5.0 & 32.6 & 3.1 & 75.3 & 2.2 &  595 \\
10.0 & 62.5 & 2.9 & 73.4 & 2.1 &  629 \\
15.0 & 86.6 & 2.7 & 67.3 & 1.9 &  608 \\
20.0 & 101.1 & 2.9 & 67.4 & 2.1 &  537 \\                                                                                                                                                                                                                                                                         \hline
\end{tabular}
\caption{Galactocentric velocities observed in our fields across the bulge for stars with [Fe/H] $>$ --1.0. }
\label{table:velpoints}
\end{table}

The line-of-sight velocity dispersion across our fields is shown in the right hand panel of Figure \ref{fig:rotation} for stars with [Fe/H] $> -1.0$. Note that no $\sigma$-clipping has been done for these measurements which are the standard deviation of all stars with [Fe/H] $> -1.0$ about their mean velocity in each field. As the major axis of the bulge/bar lies at an angle of about $20^\circ$ to the Sun-center line, fields at negative longitudes are further from the Galactic centre than fields at the same $|l|$ on the positive side. All three ranges of magnitude are included in each data point, except for our field at $(l,b) = (20^\circ, -7.5^\circ)$, which is missing the faintest stars. The velocity dispersion in the bulge is higher at lower latitude, most significantly in the central region, and the dependence of dispersion on latitude decreases with $|l|$. Across longitude, the dispersion at $b = -10^\circ$ is flatter than at $b = -5^\circ$.

The dispersion plot in the right hand panel of Figure \ref{fig:rotation} includes the data from the BRAVA survey at $b=-4^\circ, -6^\circ$ and $b=-8^\circ$ \citep{Kunder2012}. The dispersion measured from the BRAVA survey is similar to the ARGOS results. The $b=-8^\circ$ fields from BRAVA have dispersions at about the same level as the $-10^\circ$ ARGOS fields rather than the higher dispersion as measured in the ARGOS $b=-7.5^\circ$ fields. In the inner region $|l|$ $< 5^\circ$, the dispersion is about 20\% higher in the BRAVA fields at $-4^\circ$ compared to the ARGOS fields at $-5^\circ$. As discussed in paper III, there appears to be a structural change in the bulge populations in the bulge at $b < -5^\circ$ and it appears that this is most significant within $|l| < 5^\circ$.  The velocity dispersion is metallicity sensitive and is lower for the more metal-rich populations. This is examined in Section 6.3. 

\subsection{Rotation of the metal-poor stars}

\begin{figure*}
  \centering
     \epsfig{file=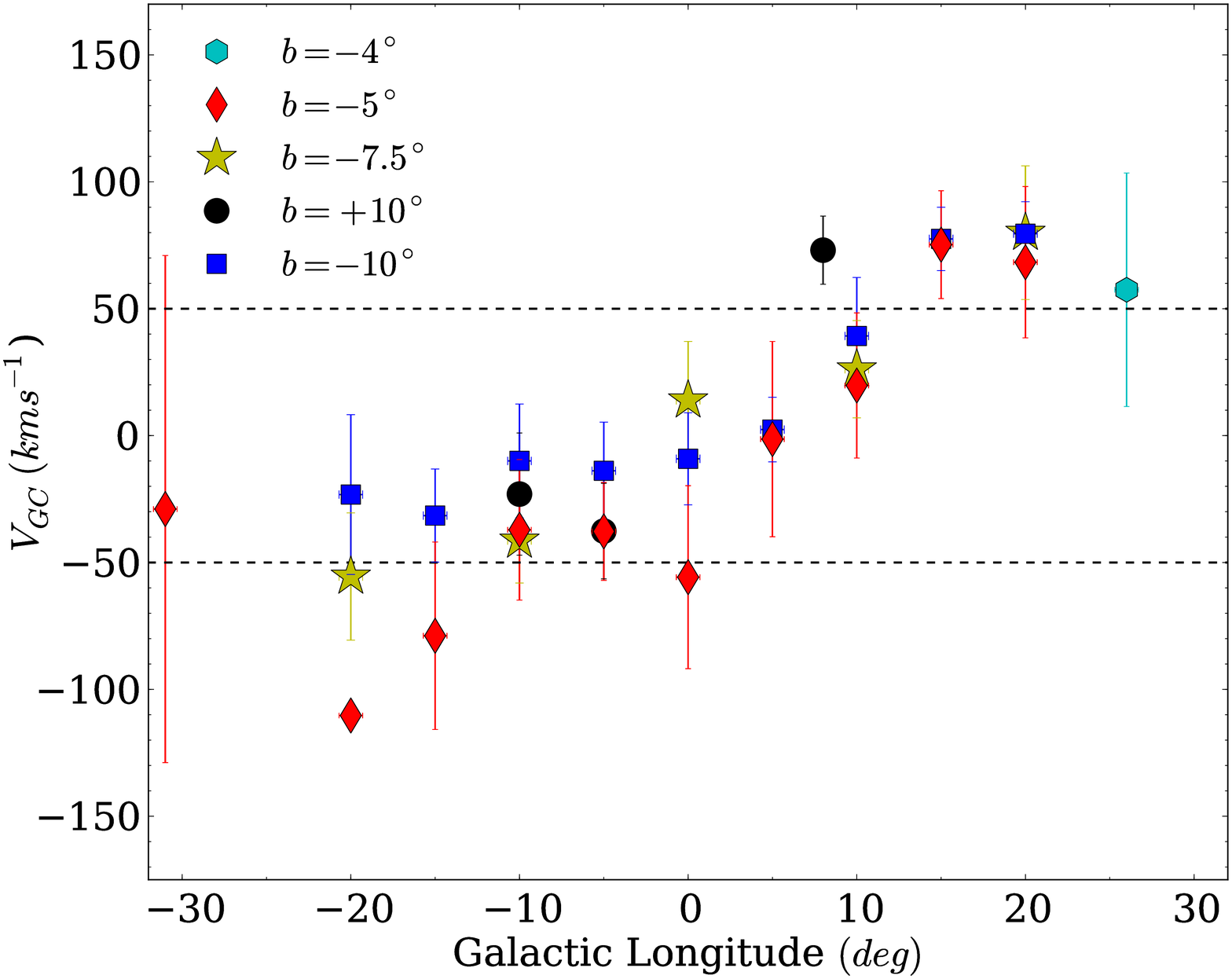,width=0.4\linewidth,clip=} 
 \epsfig{file=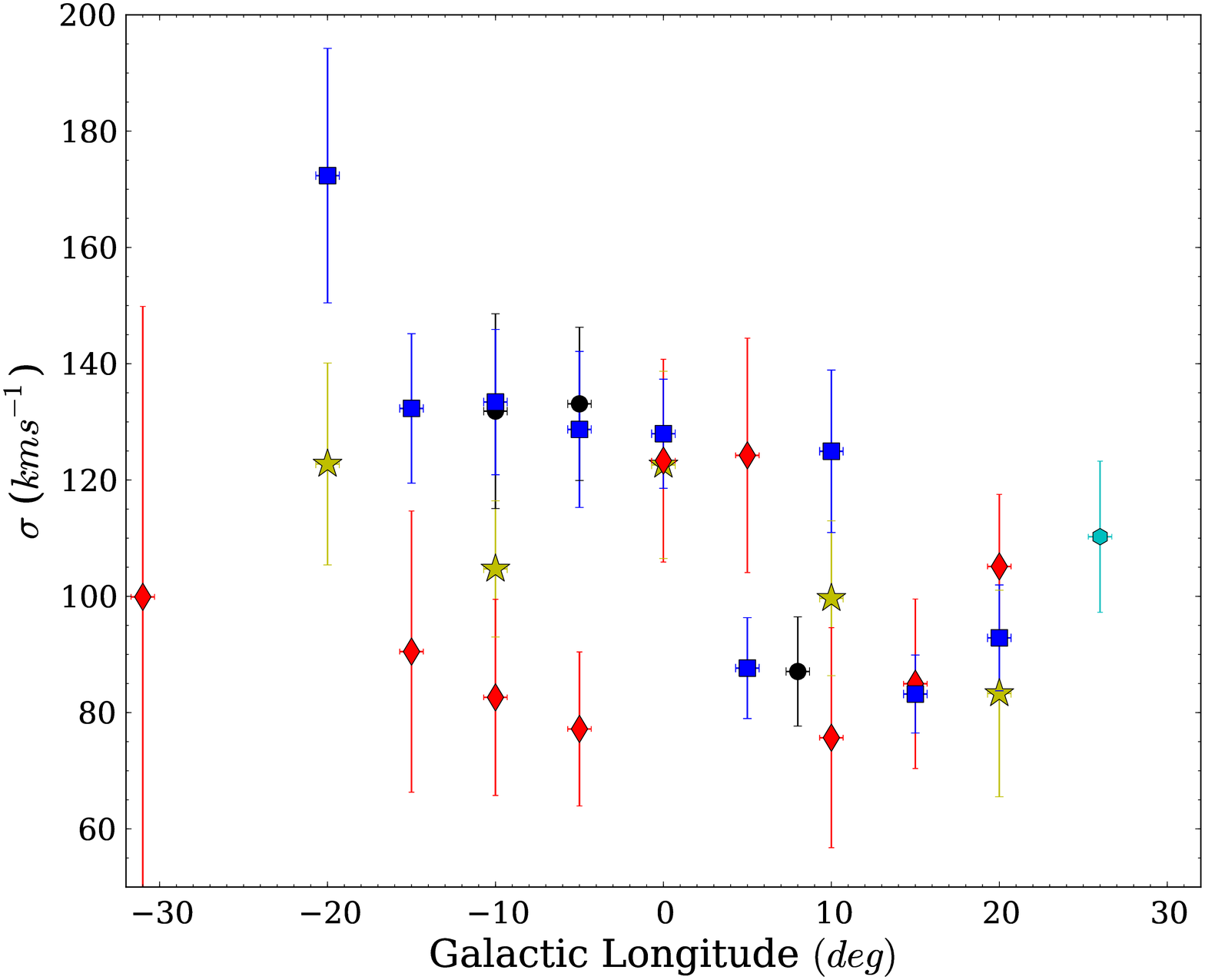,width=0.4\linewidth,clip=} \\ 
      \caption{Rotation (left) and dispersion measurements (right) for the 750 stars within $|y|$ $<$ 3.5 kpc with [Fe/H] $\le$ --1.0. The red diamonds are for $b=-5^\circ$, the yellow stars are $b=-7.5^\circ$, the blue rectangles are $b=-10^\circ$ and the black circles are for $b=+10^\circ$. The error bars in the vertical direction represent the sampling error for each field. }
  \label{fig:fehpoor}
\end{figure*}

We now proceed to investigate the rotation of the sample of stars with [Fe/H] $\le -1.0$. These stars comprise about 5\% of the data, or about 750 bulge stars within a distance $|y| <  3.5$ kpc. The rotation curve and dispersion profile for all stars with $ |y| < 3.5$ kpc and [Fe/H] $< -1.0$ are shown in the left and right panel of Figure \ref{fig:fehpoor}, respectively.

The mean rotation for the metal-poor stars is lower than for the metal-rich stars ([Fe/H] $> -1.0$), by about 50\%. The rotation in both high ($b=-10^\circ$) and low ($b=-5^\circ$) latitude fields is similar at positive longitudes on the near side of the bulge.  On the far side of the bulge the high latitude dispersion has a flatter profile. Note that the sample size in some fields is small; there are only two stars in the field at $(l,b) = (-20^\circ,-5^\circ)$. The low metallicity stars show a net rotation, and this rotation persists to the lowest metallicities  [Fe/H] $< -1.5$, but the number of stars is small and the error bars correspondingly larger (see Figure \ref{fig:counter}).  There is a larger fraction of more metal-poor stars at high latitudes than low latitudes, with about 400 stars at $b=-10^\circ$, 100 stars in the three fields at $b=+10^\circ$ and 100 stars in our fields at $b=-5^\circ$ and $b=-7.5^\circ$.  \citet{Harding1993} had earlier noted a similar change of kinematics with metallicity in a single bulge field at $(l,b) = (-10^\circ,-10^\circ)$.

The velocity dispersion of the metal-poor stars in Figure \ref{fig:fehpoor} is significantly higher than that of the metal-rich stars in Figure \ref{fig:rotation}.  Metal-poor stars in fields at a latitude of $b=-10^\circ$ (blue symbols) show a greater increase in dispersion than the fields at  $b=-5^\circ$ (red symbols), relative to stars at [Fe/H] $> -1.0$. At $b= \pm 10^\circ$, the dispersion on the far side of the bulge is higher than the near side of the bulge, shown in the left panel of Figure \ref{fig:fehpoor}. No sigma clipping has been used in the plot, because the sample is small and we did not want to exclude high velocity outliers. No high velocity outliers are seen at this latitude at positive longitudes. At negative longitudes there is a group of stars at $b=-10^\circ$ with velocities $> 300$ \kms (see the right hand panel of Figure \ref{fig:groups}). Four of the 31 stars in the field $(l,b) = (-20^\circ,-10^\circ)$, with an [Fe/H] of $-1.2$, have velocities between $300 - 330$ \kms and lie within \rgc\ $<$ 4.0 kpc of the Galactic centre. If these four stars are removed, the velocity dispersion decreases from $\sigma$ = 175 \kms\ to $\sigma$ = 125 \kms. The higher dispersion seen for more metal-poor stars is discussed further in Section 7.1. The asymmetry in the dispersion seen for the stars at $b=-10^\circ$ in Figure \ref{fig:fehpoor} also suggests there may be a relationship between the halo stars of the inner Galaxy and the bar.  In simulations which include a stellar halo component, the halo in the inner part becomes elongated and makes a so called halobar. Assuming that a non-negligible number of stars follows the dynamics of the inner part of the halo,  this would provide a metal-poor component of halo stars that may resemble the dynamics in Figure \ref{fig:fehpoor} (see \citet{Athanassoula2007}). 

\subsection{Kinematics of stars with [Fe/H] $> -1.0$}

 \begin{figure*}
  \centering
  \epsfig{file=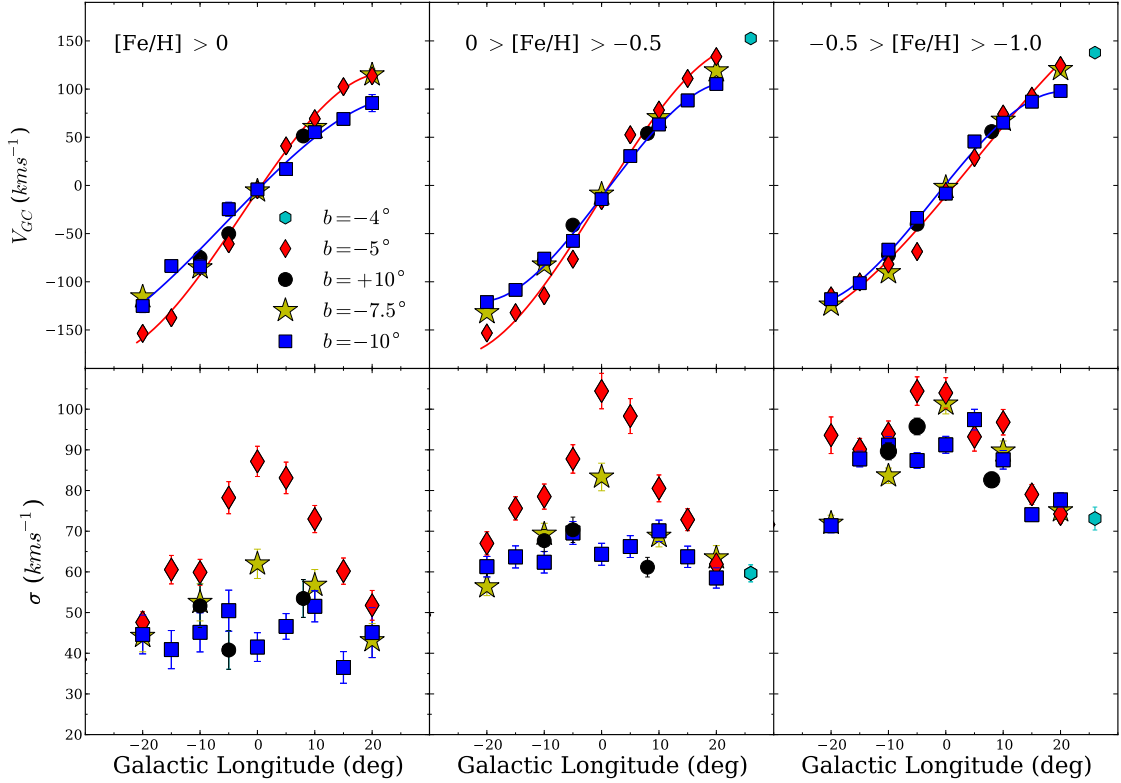,width=1.0\linewidth,clip=} \\  
     \caption{Rotation (top panel) and velocity dispersion (bottom panel) across the bulge to $|l| = 20^\circ$ for the 16,600 stars with [Fe/H] $> -1.0$ within $<$ 3.5kpc of the Galactic centre. The three plots correspond to different metallicity bins, from left to right in decreasing [Fe/H] as shown. Note that the discrete bins are used to represent stars of our components A,B and C from left to right. Although the rotation curves are similar, the dispersion clearly demonstrates the difference in kinematics of stars with [Fe/H] $> -0.5$ and with [Fe/H] $< -0.5$. There are 3100, 8600 and 4900 stars in each plot, from left to right. The red diamonds are for $b=-5^\circ$, the yellow stars are $b=-7.5^\circ$, the blue rectangles are $b=-10^\circ$ and the black circles are for $b=+10^\circ$. }
     \label{fig:ABC}
\end{figure*}

In paper II we identified stars with [Fe/H] $>$ --0.5 to be of the bulge, and we believe these stars had a common dynamical history.  We decomposed our sample of observed stars into metallicity components representative of the populations in the inner region. The most metal-rich component A with mean [Fe/H] $\approx +0.15$ is a relatively thin and centrally concentrated part of the boxy/peanut-bulge. Component B with mean [Fe/H] $\approx -0.25$ is a thicker boxy/peanut-bulge structure seen in relatively constant fraction across our latitude range in the inner region.  We speculate that the stars in component C, with mean [Fe/H] $\approx  -0.70$,  are from the early thick disk which may have been puffed up in the inner region during a bar buckling event. We note again that component C does not appear to be part of the boxy/peanut structure of the bulge.  The components are overlapping in metallicity, and component B extends down to near [Fe/H] $= -1.0$, well into the metallicity range of component C. Therefore, stars  as metal-poor as [Fe/H] $\approx -1.0$ are all contributing to the bulge in the inner region. The stars with [Fe/H] $< -1.0$ show slow rotation and high dispersion (see Figure \ref{fig:fehpoor}) and we attribute them to the metal-weak thick disk and halo.

We examine the kinematics of stars in metallicity intervals corresponding to our components A, B and C to demonstrate that A and B share common properties distinct from C. Figure \ref{fig:ABC} shows the rotation (top panels) and velocity dispersion (bottom panels) for stars in metallicity intervals  [Fe/H] $> 0$ (A), $-0.5 < $ [Fe/H] $\le 0$ (B) and $-1.0 < $ [Fe/H] $\le -0.5$ (C). The rotation for all three [Fe/H] intervals is fairly similar, with component B showing the fastest rotation, about 20\% faster than components A and C at $b=-5^\circ$.  Although the rotation curves are similar, the velocity dispersion profiles are clearly different between the three components.   The dispersion profiles for components A and B have similar morphologies with changing latitude;  component A is a kinematically colder replica of component B. We note that component A shows a relatively high dispersion at $b= \pm 10^\circ$ and $l$ = $-5^\circ, +10^\circ$ which is seen to a lesser extent also in component B. These fields correspond to similar distances along the bar of about 2.0 kpc.  The dispersion for the most metal-rich fraction of stars (A) is about 30$\%$ lower than for stars associated with component B.  The relative change of dispersion with latitude for components A and B is consistent with a lower scale height for component A than for B. 

Component C however has a clearly different dispersion profile and latitude dependence relative to stars with [Fe/H] $> -0.5$, with a relatively latitude-independent dispersion out to $|l| = 10^\circ$.  At larger longitudes, $|l| > 10^\circ$, the dispersion decreases out into the thick disk further from the Galactic centre (see the far right panel in Figure \ref{fig:ABC}). The common rotation shape and dispersion profiles seen for components A and B suggest that A and B share a similar formation history or common formation mechanism.  The similar rotation but different dispersion for component C indicates that it is a distinct population, although it is part of the central rotating bulge in the inner region.  This could be due to the different initial velocity dispersion of the thin and thick disks before the instability event. 

\subsection{Stellar streams in the bulge}

\begin{figure*}
  \centering
\vspace{-10pt}
\epsfig{file=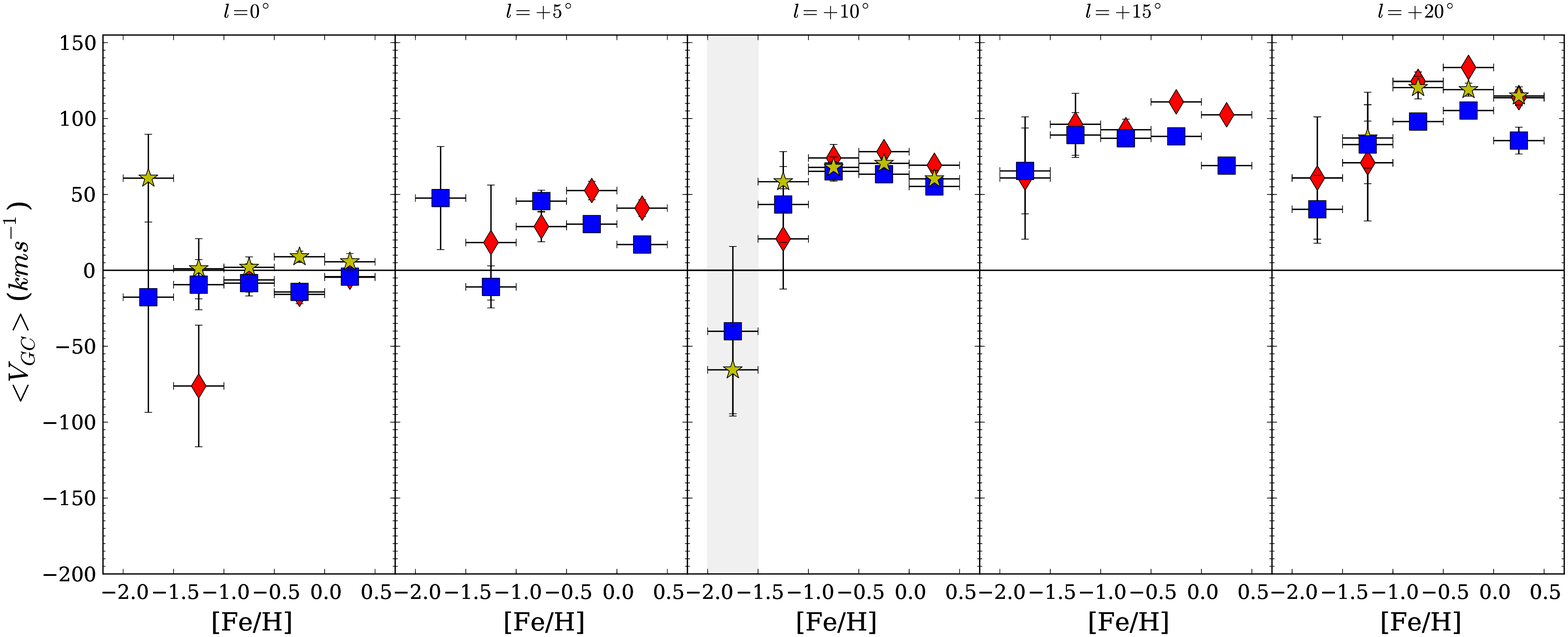,width=1.05\linewidth,clip=} \\
\vspace{-10pt}
\epsfig{file=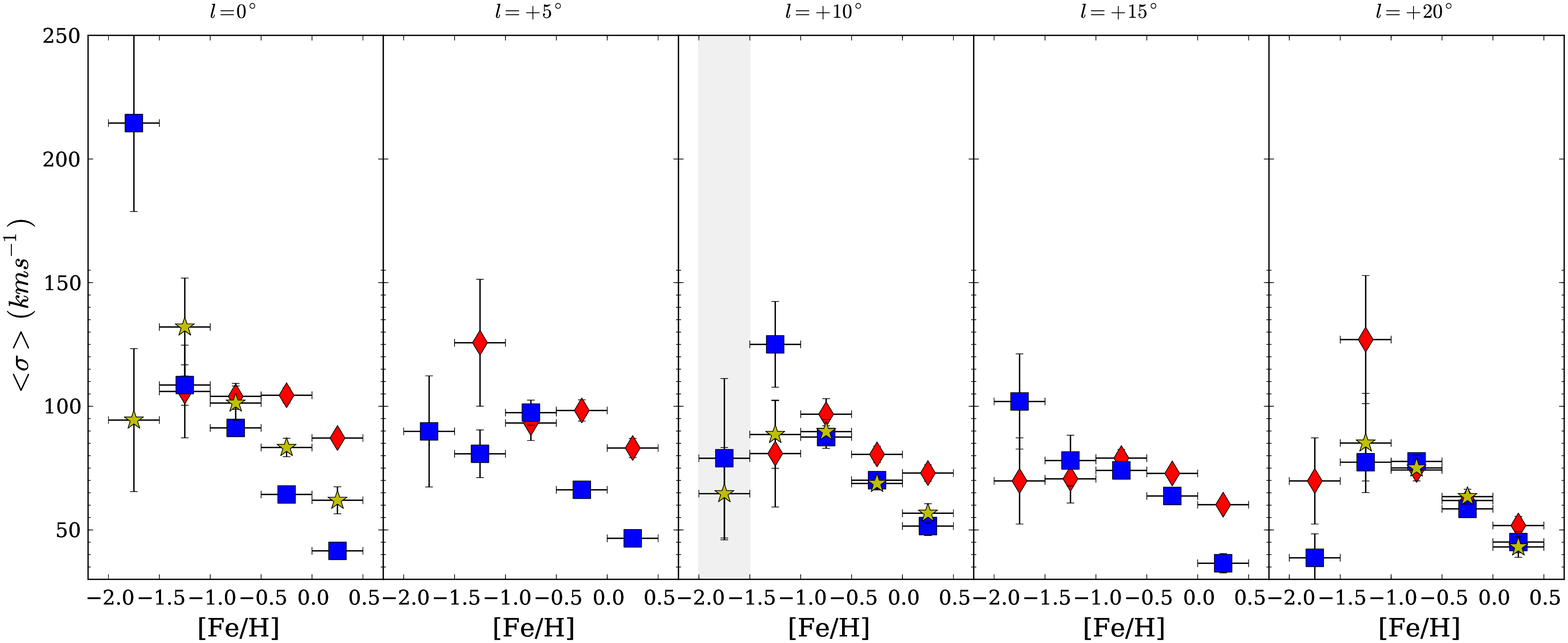,width=1.05\linewidth,clip=} \\
\caption{The rotation (top) and dispersion (bottom) versus [Fe/H] in 0.5 dex bins across positive longitudes from $l  = 0^\circ$ to $l = 20^\circ$. Diamonds are for $b=-5^\circ$, stars are $b=-7.5^\circ$, rectangles are $b=-10^\circ$. Shaded panels indicate a group of stars in the inner regions which are at large negative velocities and low metallicities.  }
\label{fig:counter}
\end{figure*}

The top panel of Figure \ref{fig:counter} shows the rotational profile of the bulge as a function of [Fe/H] in the five bins of  positive longitude, where the stars are divided into metallicity bins of 0.5 dex, except for the most metal-poor stars of which there are very few, so the lowest [Fe/H] bin takes all stars from the most metal-poor star  identified ([Fe/H] = $-2.8$) to [Fe/H] = $-1.5$. The most metal-rich bin is from [Fe/H] = 0 to 0.5 dex. These bins correspond approximately to the components which we label A-E, in order of decreasing [Fe/H], as outlined in Section 2.  

\begin{figure*}
  \centering
  \epsfig{file=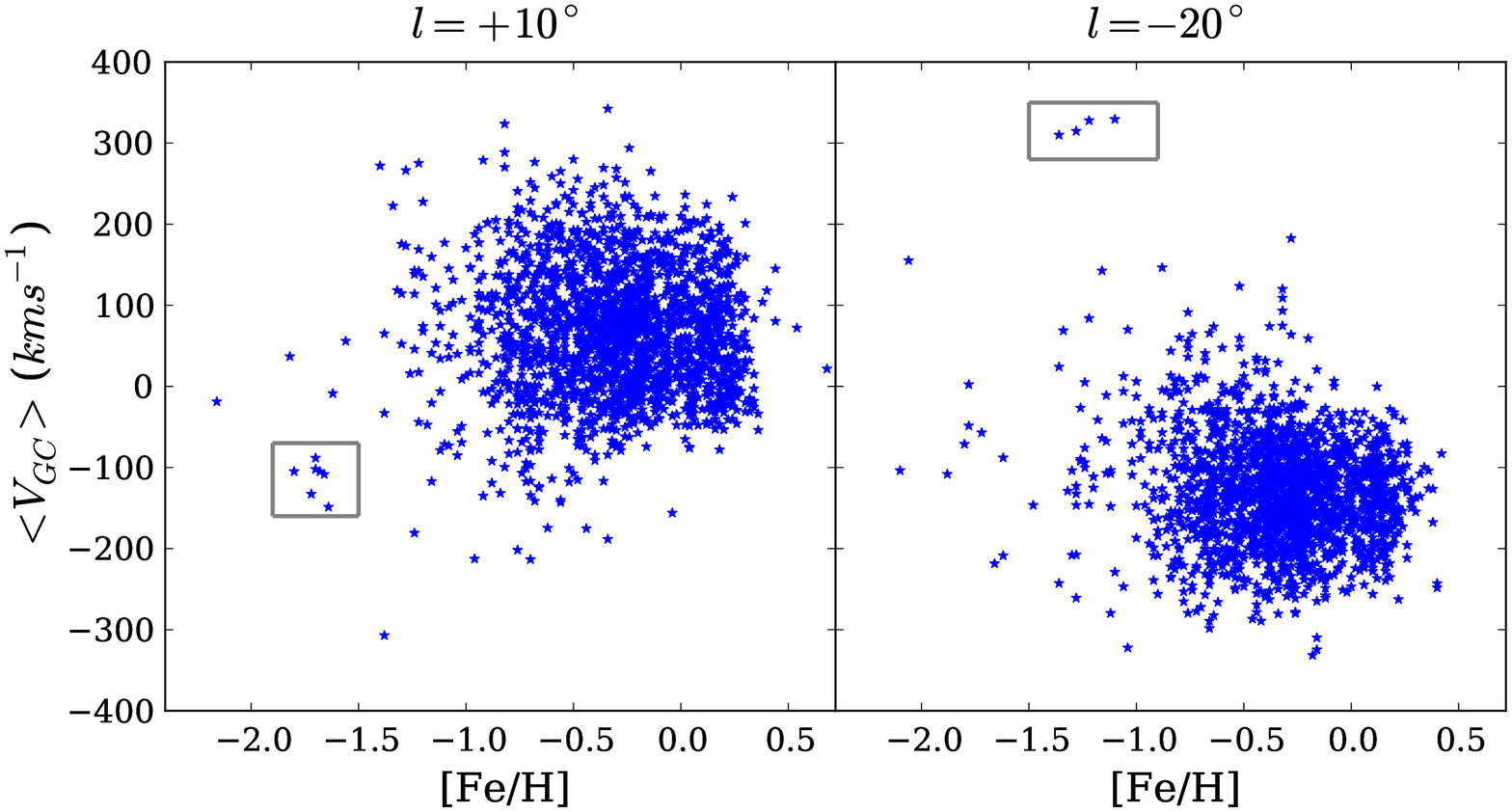,width=0.9\linewidth,clip=} \\  
     \caption{Potential metal-poor moving groups in the fields at $l=+10^\circ, -20^\circ$ that are located within \rgc\ $<$ 4.0 kpc. }
     \label{fig:groups}
\end{figure*}

\begin{figure}
  \centering
\epsfig{file=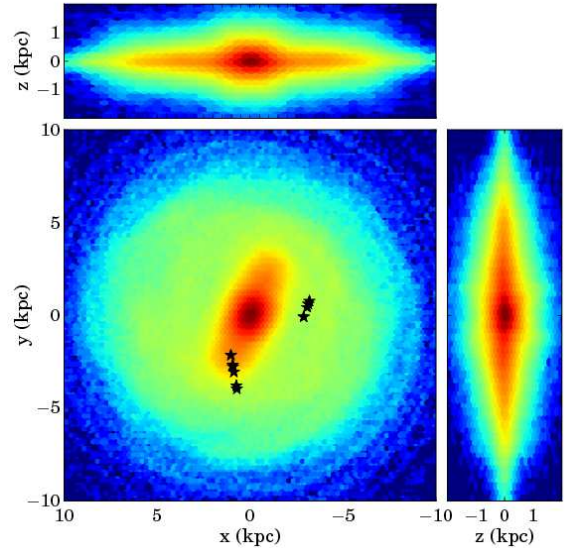,width=0.9\linewidth,clip=} 
     \caption{Surface density projections of our N-body model scaled to dimensions of the Milky Way. The Sun would be located at x=0,  y = --8 kpc and the bar is at an angle of about 20$^\circ$  to the Sun-centre line. The two groups of individual stars plotted correspond to the two groups shown in Figure \ref{fig:groups} at positive (near side of the bar) and negative (far side of the bar) longitudes, respectively.  }
     \label{fig:liamodel}
\end{figure}

The upper panel of Figure \ref{fig:counter} shows the existence of a group of metal-poor counter-rotating stars in the bulge at positive longitudes. The negative longitudes examined in this way show the same overall kinematic behaviour as a function of [Fe/H] with latitude, but do not reveal any distinct counter rotating-groups in any [Fe/H] bin. The counter-rotating stars are shown in the shaded panels in the fields $(l,b) = (10^\circ,-7.5^\circ)$ and $(l,b) = (10^\circ,-10^\circ)$ . There are seven stars in total in these fields with [Fe/H] $< -1.5$ and these are shown individually in Figure \ref{fig:groups}. Six of these stars which are shown in the left hand panel of Figure \ref{fig:groups} are also co-located in distance at around 4.9 kpc along the line of sight from the Sun, with a distance spread of $\sigma$ = 0.6 kpc (see Figure \ref{fig:liamodel}).  These six stars have a metallicity of ($\mu${([Fe/H])}, $\sigma${([Fe/H])}) = (-1.68, 0.03) and $\alpha$-enhancement range of ($\mu${([$\alpha$/Fe])}, $\sigma${([$\alpha$/Fe])}) = (0.50,0.17) and the five stars in the field at $(l,b) = (10^\circ,-7.5^\circ)$ are moving with velocities of between $-90$  and $-133$  \kms.  There is one star in this group located in the field $(l,b) = (10^\circ,-10^\circ)$, with [Fe/H] = $-1.6$ and \vgc = $-148$ \kms. This group may be an independent stellar stream in the bulge at a distance of around 4.9 kpc from the Sun, with a line of sight velocity of ($\mu(${\vgc}), $\sigma$({\vgc})) = (-115 \kms, 22 \kms). From its location it seems this group could be related to the corners of the bar.  Comparing the percentage of stars in this group by using their K-magnitude compared to the 2MASS stars in the field, about 200 stars are expected to belong to this low-metallicity group from of all stars in the 2MASS count for this field.

The right hand panel of Figure \ref{fig:groups} shows four stars at negative longitudes which are tightly clumped in distance (as noted in Section 6.2) and velocity, with ($\mu(${\vgc}), $\sigma$({\vgc})) = (320 \kms, 8 \kms). This group of stars lies in the disk and is located at about 8.4 kpc away from the Sun along the line of sight with a distance spread of $\sigma$ = 0.34 kpc (see Figure \ref{fig:liamodel}). These stars are also very similar in their metallicity, with ($\mu${([Fe/H])}, $\sigma${([Fe/H])}) = (-1.24, 0.09) and have an $\alpha$-enhancement range of ($\mu${([$\alpha$/Fe])}, $\sigma${([$\alpha$/Fe])}) = (0.38,0.15).  Comparing the percentage of stars in this group by using their K-magnitude compared to the 2MASS stars, a lower limit of about 100 stars are expected to belong to this group from all stars in the 2MASS count for this field (not taking into account 2MASS incompleteness given their mean K-magnitude of about K = 13.2).  

The panels in Figure \ref{fig:counter} span longitudes to the edge of the bulge, at $l = 0^\circ$, $l=5^\circ$ and $l = 10^\circ$ and also our fields outside the bulge out into the disk at $l = 15^\circ$ and $l = 20^\circ$. Three latitudes zones are included in these plots, at $b=-5^\circ,-7.5^\circ$ and $-10^\circ$. Missing points in the plots indicate no stars in this [Fe/H] bin for the corresponding ($l,b$).  From these rotation plots as a function of [Fe/H], it is clear that the thicker boxy/peanut bulge component B, corresponding to 0 $>$ [Fe/H] $>$ --0.5 is generally the fastest rotating. This component rotates most rapidly in the lower latitude fields, $b=-5^\circ$. The rotation speed increases with longitude. The increase is more rapid for the metal-rich components, at around $48$  \kms kpc$^{-1}$ for [Fe/H] $> -1.0$, compared to about  $27$  \kms\ kpc$^{-1}$  for the metal-poor stars with [Fe/H] $< -1.0$. 

All stars with [Fe/H] $> -1.0$ (components A,B,C) have fairly consistent relative rotation trends across longitude, and components A and B show the largest difference in rotation velocity with latitude near the corners of the bulge, at $l=15^\circ$. The trends seen with longitude in the more metal-poor stars ([Fe/H] $< -1.0$) are not as consistent as for the more metal-rich ([Fe/H]$ > -1.0$) stars. Component D rotates more slowly than A,B,C; the rotation of component E depends on the field. 

From the bottom panel of Figure \ref{fig:counter}, it is clear that the velocity dispersion mostly increases as metallicity decreases down to  [Fe/H] $< -1.0$ ;  for lower metallicities it can decrease or increase depending on the field. Note that in the field where the maximum dispersion is found, $\sigma \approx$ 215 \kms\ at $ (l,b)  = (0^\circ,-10^\circ)$,  there are only 9 stars in this bin of [Fe/H], one of which has a \vgc\ = 412 \kms\ (there is no sigma clipping:  all stars are included in the plot). The error bars become larger across all longitudes for the metal-poor stars,  [Fe/H] $< -1.0$, as the number of metal-poor stars is small.

\section{Comparisons to models} 

In this section we compare our data to a series
of six simulations of a rotating disk embedded in a live dark matter
halo. None of our models contains gas, i.e. there is no star formation
and no information on metallicity for us to compare to. We also
include only one disk component, instead of two separate thin and
thick disk components. The disk and the halo initial density distributions are as described
in \citet{Athanassoula2007} and have been built in equilibrium
in each other's potential as described in that paper. All models have
identical initial conditions for the disk, with a Toomre
parameter equal to Q=1.2 and a vertical height of 0.2 times the disc
scale length. In this way some effect from the thick disk is
included. We consider halo core sizes in the range
between 0.1 and 4 times the initial disc scale length. The former
leads to model with a submaximum disk, while in the latter the disc is
maximum (for a discussion on the effects of maximum and sub-maximum
disks on bar growth and evolution, see \citet{Athanassoula2002b}. Their evolution was followed using the Gyrfalcon N-body
code \citep{Dehnen00, Dehnen02}.

All models give rotation curves that compare reasonably well
with the data, but the changes in velocity dispersion across longitude
and offsets between $b = -5^\circ$ and $b =-10^\circ$ are model-dependent. 
Although the kinematic profiles of these models are similar to those observed, none
fits across all latitudes. The variation of velocity dispersion across
longitude is sensitive to the bar properties. For bars that
are too short with respect to the observations, the dispersion decreases more 
rapidly than observed at the highest latitude and across the longitude range $|l| < 20^\circ$. The dispersion
profile at the lowest latitude is sensitive to the length of the
bar which offers a complementary test of the observed scale length
of the bulge. 

\begin{figure}
  \centering
  \epsfig{file=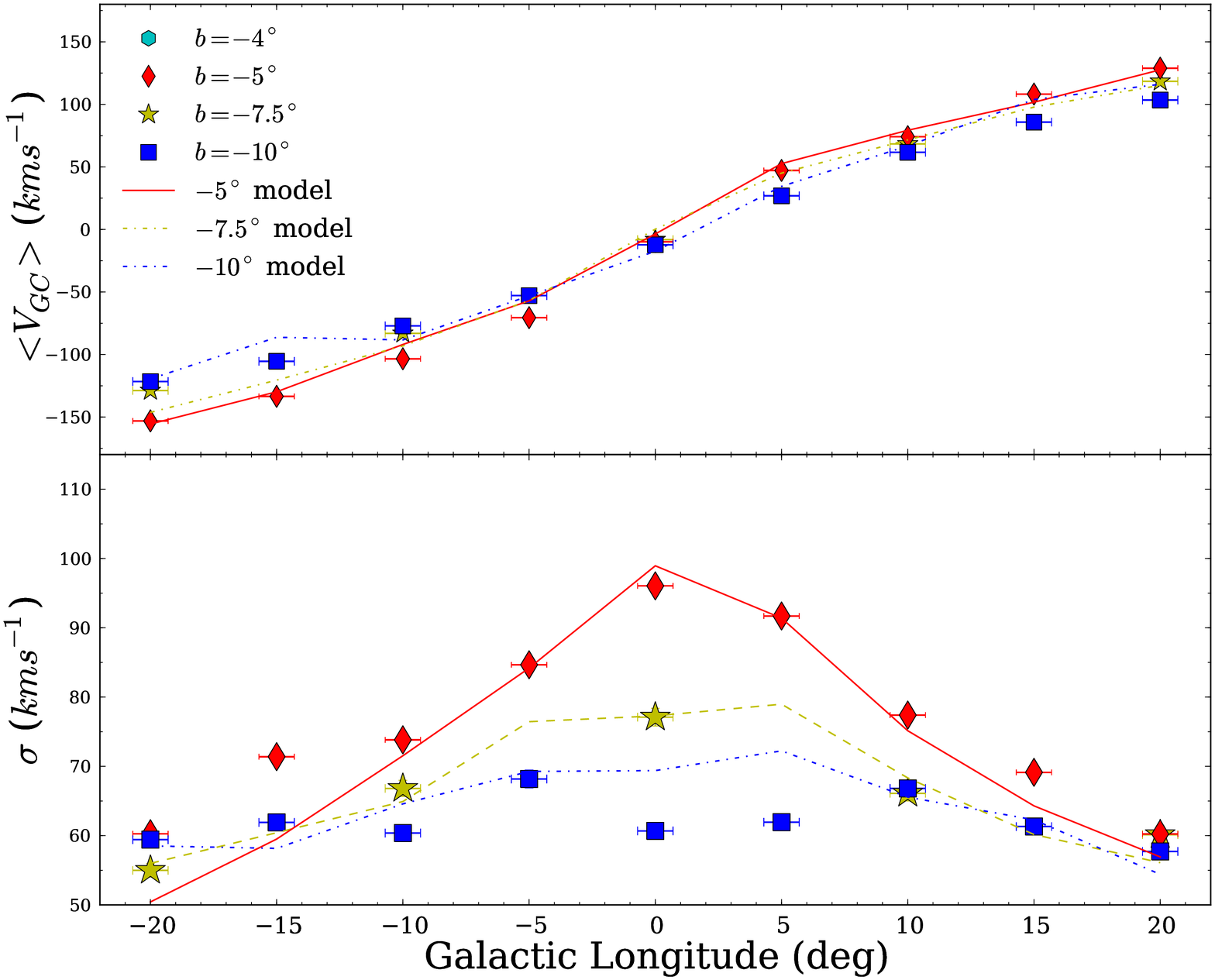,width=0.95\linewidth,clip=} 
     \caption{The rotation and dispersion of our model (lines) plotted with our data (points) for stars with [Fe/H] $>$ --0.5. The red diamonds are for $b=-5^\circ$, the yellow stars are for $b=-7.5^\circ$ and the blue rectangles for $b=-10^\circ$. The data errors in the vertical direction are smaller than the symbols and the error bars in the horizontal direction represent our two-degree fields).}
       \label{fig:liarotationAB}
\end{figure}

As already discussed in Section 2, we have associated stars with [Fe/H] $>$ --0.5 with the boxy/peanut bulge. Our model shows the X-shape structure in the density distribution of the stars at high latitudes in the inner region, showing a very good match between observational data for stars with [Fe/H] $>$ --0.5 at the minor axis, similarly to the model discussed in \citet{Ness2012}. Figure \ref{fig:liarotationAB} compares the kinematic data for the stars with [Fe/H] $>$ --0.5 to the model. The model velocities have been scaled in amplitude to best match the dispersion at $b = -7.5^\circ$ out to $|l| = 20^\circ$. The rotation curves of the model (top panel of Figure \ref{fig:liarotationAB}) are a reasonable fit to the data for the stars with [Fe/H] $>$ --0.5. In both the model and observations, at $l = \pm$  $20^\circ$ the rotation velocity at $b=-5^\circ$ is approaching 150 \kms\ and at $b=-10^\circ$ it is about 120 \kms. The rotation velocity is marginally high in the model compared to data at positive longitudes at $b=-10^\circ$.  In the model, the Sun is located at 8 kpc from the Galactic centre along the line joining the Sun to the centre of the Galaxy and the bar is pointing 25 degrees into the first quadrant.  Adjusting the bar angle by $\pm 10^\circ$ relative to the Sun-centre line changes the gradient of the rotation by about 15\%: larger bar angles give flatter rotation curves. Although the dispersion profile is model dependent, the characteristics of the profile across $(l,b)$ are generic to our N-body models of instability-generated boxy/bulges.

From the bottom panel of Figure \ref{fig:liarotationAB}, it is clear that the overall shape of the dispersion profile in the model matches well the observations across our fields for our more metal-rich stars (components A and B, see Figure \ref{fig:ABC}). At $b=-5^\circ$, the observed and model dispersions are a fairly good match except at $l < -10^\circ$ where the model dispersion decreases faster than the observations. At $b=-10^\circ$, the observed and model dispersion are both fairly flat across the longitude range but in the inner region the observations are about 15$\%$ lower than the model. The gradient of the model and data are, however, similar at the three latitudes, and at $b=-5^\circ$ both model and data show an asymmetry at $l = \pm$ $5^\circ$, where the dispersion at positive longitudes is higher than the dispersion at the negative longitudes. The feature in the dispersion seen near the edges of the bar at $l=-5^\circ$ and $l=-10^\circ$ at the higher latitude of $b = -10^\circ$ (see also Figure \ref{fig:ABC}), is not reproduced in the model. 

\begin{figure}
  \centering
   \epsfig{file=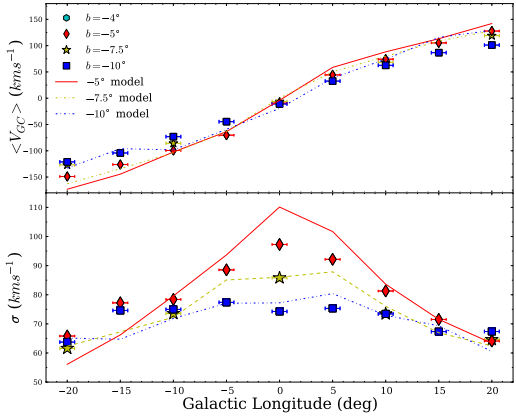,width=0.95\linewidth,clip=} 
     \caption{The rotation and dispersion of our model (lines) plotted with our data (points) for stars with [Fe/H] $>$ --1.0. The red diamonds are for $b=-5^\circ$, the yellow stars are for $b=-7.5^\circ$ and the blue rectangles for $b=-10^\circ$. (The data errors in the vertical direction are smaller than the symbols and the error bars in the horizontal direction represent our two-degree fields.) Note the scaling from code units to \kms\ is different to that of Figure \ref{fig:liarotationAB}.}
       \label{fig:liarotation}
\end{figure}

In Figure \ref{fig:liarotation} we compare the model to stars with [Fe/H] $>$ --1.0. Although the dispersion of stars in our component C does not reflect that of the model and is not involved in the boxy/peanut structure \citep{Ness2012}, we include this comparison because this component is still cylindrically rotating similarly to A and B. It is possible that there are a smaller number of stars within the model contributing to this dispersion distribution seen in C. We cannot break up our model in the same way we do the data because the model has no metallicity or age information. Furthermore the components are overlapping; stars that we identify with the boxy/peanut bulge (components A and B) do extend in the tail down to [Fe/H] $\approx$ --1.0 (see Figure \ref{fig:components}). The single disk model is, however, unlikely to include the thick disk stars that we associate with component C in the same proportion as in the Galaxy. In Figure \ref{fig:liarotation}, the observed and model velocities and dispersions are a fairly good match, although the rotation of the model is faster than the rotation of the stars and the dispersion of the model in the central regions at $b=-5^\circ$ is about 15$\%$ larger than the data. At higher latitudes, the dispersion flattens in both the model and data although the dispersion of the model is about 10$\%$ too low at $|l| > 10^\circ$. The model has again been scaled in amplitude to best match the dispersion at $b=-7.5^\circ$ out to $|l| = 20^\circ$. Note that the velocity scaling factors for the model used in Figures \ref{fig:liarotationAB} and \ref{fig:liarotation} are different, and were adjusted to best fit each data set for stars [Fe/H] $>$ --0.5 and stars [Fe/H] $>$ --1.0, respectively. A larger scaling factor for the velocity code units is used to best match the dispersion for stars with [Fe/H] $>$ --1.0. The bar is rotated at about 25$^\circ$ to best fit the dispersion profile for both Figures.

As already mentioned, our model has a single disk and no star formation and this could explain why its final bulge profile  does not reflect all the populations we see in the central region of the Galaxy. Models that have two initial disks may offer more insight. If component C, the thick disk, is a separate structure, it may influence the dynamical instability mapping of stars from the original disk into the bulge without itself being mapped into the X-shaped structure. Furthermore, given that we find two distinct components A and B for the bulge, and associate A with the colder thin disk in the inner region, it may not be surprising that we can not find an exact match between model and data. The origin of the duality we see in the bulge, with a kinematically colder metal-rich component and kinematically hotter more metal-poor component extending to higher latitudes, is not clear. It may be due to the different redistribution of colder higher metallicity disk stars and lower metallicity disk stars with higher orbital energies (as seen in the solar neighbourhood) to larger heights from the plane during the instability event in a single disk scenario. Alternatively it may be a consequence of the instability of two disks or else a second star formation event during the instability process close to the plane. Additionally, although the model was scaled to the light profile seen in the COBE image, the relative scales and densities of the components of thin disk, bulge and thick disk in the bulge regions will affect the dispersion profile. It seems unlikely that a simple single-disk model in which all these stars have formed before the instability event will match exactly over all $(l,b)$.  Adjusting the angle of the bulge with respect to the Sun within a wider estimated range of $15^\circ$ to $30^\circ$ \citep{Stanek1997, Binney1997, Bissantz2002} did not significantly improve the match.  Further work will be done using N-body models and adjusting the input parameters to fit the observational data to better constrain the models.

 \section{Discussion}

In Section 7 we briefly described the comparison of our data to a sequence of models with different halo core radii: the best fits were found for models with large core radii. These will have maximum disk rotation curves,
while those with the small halo core radii will be initially clearly
sub-maximum and will evolve towards maximum with time
\citep[see][for a review]{Athanassoula2012}. Unfortunately, this
finding does not allow us to set  
constraints on the dark matter distribution in the central parts of
our Galaxy, because our models do not include a gaseous component. As
discussed in \citet{AthanassoulaMR12}, adding such a component  
will lead to a less strong bar and thus to a different quality fit. In
other words, the properties of the bar do not depend only on the halo
mass distribution, but also on the gas, its evolution and its
properties. We thus cannot set any set any strong constrains on the
former without having a reasonably solid knowledge of the latter,
including its evolution in the past.

\begin{table}
 \centering
  \begin{tabular}{| p{0.6cm} | p{1.5cm} | p{1.5cm}| p{1.7cm} | p{1.5cm} | }
 \hline
 Name&  \multicolumn{2}{|c|}{\vgc\  $kms^{-1}$} &   \multicolumn{2}{|c|}{$\sigma$  $kms^{-1}$}  \\
 \hline
& l = $\pm 5^\circ$ & $ l = \pm  10^\circ$ &  $ l = \pm 5^\circ$ & $ l = \pm 10^\circ$ \\
\hline
&  \multicolumn{4}{|c|}{$b=-5^\circ$ }  \\
 \hline

A & 51.6 $\pm$ 5.5 &  77.8 $\pm$ 4.6 &  80.5$\pm$ 3.9 & 65.5  $\pm$ 3.2 \\
B & 63.5 $\pm$ 5.6 & 95.6 $\pm$ 3.9 & 93.5 $\pm$ 3.9 & 79.6 $\pm$ 3.2 \\
C & 46.3 $\pm$ 10.0 & 78.0 $\pm$ 8.9 & 98.2 $\pm$ 7.0 & 95.3  $\pm$6.3 \\
\hline
&   \multicolumn{4}{|c|}{$b=-10^\circ$}  \\
\hline
A & 22.3 $\pm$ 6.5 & 74.6 $\pm$ 6.4 &  49.2 $\pm$ 4.5 & 47.2 $\pm$ 4.7 \\
B & 44.0 $\pm$ 3.9 & 70.4$\pm$ 3.7 & 67.9$\pm$ 2.8 & 65.8$\pm$ 2.6 \\
C & 40.6 $\pm$ 6.3 & 65.9 $\pm$ 6.3 & 93.2 $\pm$ 4.6 & 89.2 $\pm$ 4.4 \\
\hline
\end{tabular}
  \caption{Rotation velocity and dispersion for different components in the bulge at $b=-5^\circ$, $ b = -10^\circ$ and $l = \pm 5^\circ$, $l = \pm 10^\circ$.}
\label{table:fehdisp}
\end{table}

We see from the ARGOS data that the boxy/peanut bulge population B is mostly the fastest rotating of our bulge components (Table \ref{table:fehdisp}). We associate the metal-rich component A chemically with the younger more metal-rich stars of the inner thin disk, and its rotation and dispersion is typically $\sim$ 80\% that of component B. This difference in rotation between A and B is a potentially important signature that we can look for in chemodynamical or age-tagged N-body models of the instability event. In the solar neighbourhood,  the velocity dispersion of the thin disk is relatively small.
However, the thin disk dispersion decreases with Galactic radius \citep{LewisFreeman1989} and all components, including the most metal-rich boxy/peanut-bulge population A that we associate with the thin disk, are kinematically hot in the inner regions.  

We now discuss the distribution of velocity dispersion for the components of the bulge.  At $b=-10^\circ$, stars with [Fe/H] $> -1.0$ show a fairly flat distribution of velocity dispersion with longitude for $|l| < 20^\circ$.  The velocity dispersion of components A and B shows structure  at $b=-10^\circ$ at longitudes of  $-5^\circ/+10^\circ$ degrees (see Figure \ref{fig:ABC}) which correspond approximately to equal distances of about 2 kpc along the inclined bulge. The velocity dispersion is higher in these fields, particularly for component A.  This structure, which is not seen in the more metal-poor component C, must correspond to some feature of the orbit structure in the bar/bulge, but it is not apparent in our N-body models and we cannot at this stage identify its nature. The Sagittarius dwarf stream is present in this direction, but it is unlikely that we are including members of this stream in our sample, because we select stars within $y= \pm 3.5$ kpc of the centre of the Galaxy. However, the effect of any contamination from the Sagittarius dwarf stream would be to cause a fake high dispersion in some fields, particularly at $|l|=5^\circ$ , $|b|=10^\circ$. Note all the dispersions are calculated from the standard deviation of all stars in the sample, without $\sigma$-clipping.

For stars with [Fe/H] $> -0.5$ at $b=-5^\circ$, the velocity dispersion decreases almost linearly with longitude at  about $-12$ \kms\ kpc$^{-1}$.    For comparison, \citet{LewisFreeman1989} measured the radial component of the velocity dispersion for the stars of the old disk of the Milky Way, from \rgc\ $= 18$ kpc to about $3$ kpc, and also the azimuthal component $\sigma_{\phi}$ of the dispersion over a more restricted range in \rgc.  In the inner region, they find that the dispersion component  $\sigma_\phi$ decreases with increasing radius at  about $-12$ \kms\ kpc$^{-1}$, very similar to the gradient in dispersion observed for our components A and B.   At the value of \rgc\ corresponding to $l = 20^\circ$,  \citet{LewisFreeman1989} find that  $\sigma_\phi$ is about $75$ \kms,  a little higher than observed for components A and B (see Figure 5).  We note that the  \citet{LewisFreeman1989} field is at a latitude of $b=-4^\circ$, closer to the plane than our $b=-5^\circ$ fields.

In our $b=-5^\circ$ fields, our observed stars are predominantly boxy/peanut-bulge members (A and B) with a smaller fraction of thick disk stars (C). The velocity dispersion gradient measured from our survey for these components A and B is in very good agreement with the results of \citet{LewisFreeman1989} for the old disk. The similar gradients for A and B suggest these populations are dynamically associated, and this would be consistent with bulge formation out of the disk. We see that component A has slower rotation and smaller velocity dispersion than component B. From Jeans' equation, that means that the scale length of A is shorter than for component B. We also argued earlier that A is more concentrated to the Galactic plane, so it has a smaller scale height than B. It seems that A is a more compact version of B, in both vertical height and radius. For component C, the observed decrease in velocity dispersion with longitude is much slower (about $-7$ \kms\ kpc$^{-1}$) at both $b=-5^\circ$ and $b=-10^\circ$, from $l = 0^\circ$ to $20^\circ$, corresponding approximately to  \rgc\ $= 0$ to $3.0$ kpc.  We associate component C with the thick disk.  The different structure and kinematical properties of component C suggest that its dynamical history has been different from that of boxy/peanut bulge itself.   

There is another scenario for interpreting components A,B
and C. Using an N-body model, it is possible to show that an
underlying classical bulge can absorb angular momentum from a forming
bar \citep{Athanassoula2003}. This will transform an initially small
isotropic non-rotating classical bulge into a triaxial and
cylindrically rotating object \citep{Saha2012}. This
object is then rotating along with the bulge formed from the disk
via the dynamical instability. This transformed structure could be seen as one of our metallicity components. Could this structure be our component B or C ?  This depends on the detailed predictions of the model: is the transformed classical bulge predicted to show the peanut structure associated with a split in the red clump stars, and how does its expected velocity dispersion compare to that of the instability-generated bulge? Given the characteristic dispersion seen in B, however, it seems more likely that component C would be a candidate for this scenario. We cannot discuss this scenario further without more predictions from the simulations. An expansion of the \citet{Saha2012} results to evaluate the kinematics and spatial distribution of the inner bulge and the transformed classical component would enable further investigation of this formation scenario in the light of our data.

The rotation curves for the metal-poor stars with [Fe/H] $\le -1.0 $ are typical of a slowly rotating population and are probably part of the halo and metal-weak thick disk populations. Their velocity dispersions are typically $100 - 150$ \kms, consistent with identification as halo stars.  

We find a kinematically distinct population with [Fe/H] $<$ --1.0 which is not aligned with the main body of rotation of the stars in the bulge region. These stars are located in our fields at $(l,b) = (10^\circ,-7.5^\circ)$ and $(l,b) = (10^\circ,-10^\circ)$ and have mean metallicities $\approx -1.7$. These stars may belong to a stellar stream associated with an accretion event of a small system by the Galaxy. There is another possibility for the origin of the very metal-poor stars observed in the inner Galaxy in our survey. According to $\Lambda$CDM simulations, the bulge region between longitudes of $10^\circ $ to $20^\circ$ is expected to contain the oldest stars in the Galaxy, likely only a few generations older than the first stars \citep{Tumlinson2010}. These stars are not necessarily the most metal-poor stars in the Galaxy but they are formed at early times in the high density pre-galactic fragments that are rapidly chemically enriched due to a fast star formation rate and are subsequently accreted by the assembling Galaxy. Stars with [Fe/H] $< -1.0$ \ also include the potential first generation star candidates now located in the bulge \citep{White2000, Diemand2005, Tumlinson2010}. These stars are expected to show distinct chemical markers \citep{Kobayashi2011} and be on more tightly bound orbits than ordinary stars of the halo \citep{Tumlinson2010}. These stars may now be associated in metallicity but are unlikely to be associated in kinematics due to the short mixing timescales in the central bulge. More detailed chemical abundance analysis will reveal more clues as to the origin of the group of seven metal-poor stars with similar radial velocities and metallicites in our fields at $(l,b) = (10^\circ,-7.5^\circ)$ and $(l,b) = (10^\circ,-10^\circ)$. It seems likely however, that these stars may belong to more conventional stellar streams associated with small systems accreted by the Galaxy (see Figure \ref{fig:groups}). A more detailed chemical analysis of all of our stars with [Fe/H] $<$ --1.0 will similarly test if these stars are all consistent with halo stars in the inner region, with comparable properties to the halo stars observed near the Sun, or if chemical markers indicate they belong to the first generation of stars predicted by $\Lambda $CDM simulations. 

Comparison of the stellar kinematics  as a function of the metallicity across longitude at $b=-5^\circ$ and $b= -10^\circ$ shows that stars with [Fe/H] $> -1.0$ have similar kinematical properties, which are dissimilar from the rotation of stars with  [Fe/H] $< -1.0$. \citet{Harding1993} and \cite{Minniti1996} reported such a kinematic break at [Fe/H] $\approx -1.0$ from surveys of K giants in the bulge.  They also found an increasing dispersion with decreasing [Fe/H] in their bulge fields.  We find this trend of decreasing dispersion with metallicity from the bulge out into the disk, for stars with [Fe/H] $> -1.0$. We also see  that the  dispersion shows a discontinuity at [Fe/H] $= -1.0$, where both the rotation and dispersion become a changing function of longitude and latitude. We conclude that stars with [Fe/H] $> -1.0$ share a common history, showing similar overall kinematics and a smooth transition out into the disk.   

Although stars with [Fe/H] $> -1.0$ are rotating cylindrically and are kinematically distinct from stars with [Fe/H] $< -1.0$, the stars with [Fe/H] $> -0.5$ have a clear kinematical identity.  The stars in our components A and B show the same dispersion trends across ($l,b$), and these are not seen for the stars with $-0.5 >$ [Fe/H] $> -1.0$ in component C (see Figure \ref{fig:ABC}).  The stars with [Fe/H] $> -0.5$ (components A and B) appear to have experienced similar dynamical processes which were not shared by the stars in component C. We will be building a model of the bulge/bar and installing it into the Galactic modelling tool Galaxia \citep{Sharma2011}. Then we can subtract the model from the Milky Way data and look for substructure.

\section{Conclusion}

Simulations of bar formation predict that the Galactic bulge is just the bar seen edge-on and not far from end-on, puffed up into a boxy morphology in the inner regions. We find cylindrical rotation for stars with [Fe/H] $> -1.0$  which transitions smoothly out into the disk. Across our three latitude zones, this rotation compares well with that for our N-body model  of a boxy bulge formed via dynamical disk instabilities. That our N-body model does not replicate the velocity dispersion measured across all latitudes may reflect that the initial inner Galaxy probably comprised multiple populations with distinct kinematics and density gradients, while our simple  model evolved from a single thin disk. 

Based on the relationship between kinematics and [Fe/H], we have argued that stars with [Fe/H] $> -1.0$ share a common origin, distinct from metal-poor stars with [Fe/H] $< -1.0$. The characteristic dispersion profiles seen for stars with [Fe/H] $> -0.5$ suggest that these stars have shared a common dynamical history. The distinct dispersion profiles for stars with [Fe/H] $> -0.5$ are compatible with the split stellar magnitude distribution of the red clump stars in the inner region which is only seen for stars with [Fe/H] $> -0.5$. The metal-poor stars with [Fe/H] $< -1.0$ show a more spheroidal rotation profile although this is not necessarily evidence of an underlying merger-generated bulge component. We propose that these stars are probably part of the halo and the metal-weak thick disk. We find kinematically distinct metal-poor groups of stars in our sample. These stars may belong to an independent counter-rotating stream or moving group in the inner Galaxy. 

We are presently working on more complete simulations, incorporating
chemical evolution and extra components, such as a gaseous disk, a
classical bulge and a thick disk. Our observation that the most metal-rich component A is rotating more slowly than component B will provide an interesting constraint on these more complete evolutionary chemodynamic models.
One of the challenges for future modelling is to find a way to estimate or constrain the fraction of an underlying classical bulge component in the Milky Way.  We have seen from the simulations  of \citet{Saha2012} that this can no longer be done reliably from the stellar kinematics alone.

\section*{Acknowledgments}

We thank the Australian Astronomical Observatory, who have made this project possible. MN, KF and EA thank the Aspen Center for Physics for their hospitality during the workshop `The Galactic Bulge and Bar' in 2011 and the NSF for partial
financial support under Grant No. 1066293.

This publication makes use of data products from the Two Micron All Sky Survey, which is a joint project of the University of Massachusetts and the Infrared Processing and Analysis Center/California Institute of Technology, funded by the National Aeronautics and Space Administration and the National Science Foundation.

This work has been supported by the RSAA and Australian Research Council grant DP098875. MN would also like to thank the Zonta organisation of Canberra for their financial support. J.B-H is supported by an ARC Federation Fellowship. G.F.L thanks the Australian research council for support through his Future Fellowship (FT100100268) and Discovery Project (DP110100678). E.A. gratefully acknowledges financial support by the CNES and by the European Commission through the DAGAL Network (PITN-GA-2011-289313). M.A acknowledges funding from the Australian Research Council through a Laureate Fellowship (FL110100012).R.R.L acknowledges financial support from FONDECYT, project No. 3130403. L.L.K is supported by the Lend\"ulet program of the Hungarian Academy of Sciences,  the Hungarian OTKA Grants K76816, MB08C 81013 and K83790 and the European CommunityÕs Seventh Framework Programme (FP7/2007- 2013) under grant agreement no. 269194. We thank the referee for the useful comments which improved the paper.

\end{document}